\documentclass[preprint,amsmath,amssymb,aps,showkeys]{revtex4-2}
\usepackage[english]{babel}
\usepackage{graphicx}
\usepackage{graphics}
\usepackage{amsmath}
\usepackage{dcolumn}
\usepackage{amssymb}
\usepackage{bm}

\begin{document}
\title{Missing Aharonov-Casher geometric quantum phase}
\author{K. Bakke}
\email[E-mail address: ]{kbakke@fisica.ufpb.br}
\homepage[Orcid: ]{https://orcid.org/0000-0002-9038-863X}
\affiliation{Departamento de F\'isica, Universidade Federal da Para\'iba, Caixa Postal 5008, 58051-900, Jo\~ao Pessoa, PB, Brazil.}

\author{C. Furtado}
\email[E-mail address: ]{furtado@fisica.ufpb.br}
\homepage[Orcid: ]{https://orcid.org/0000-0002-3455-4285}
\affiliation{Departamento de F\'isica, Universidade Federal da Para\'iba, Caixa Postal 5008, 58051-900, Jo\~ao Pessoa, PB, Brazil.}

\begin{abstract}

From the interaction of the permanent magnetic dipole moment of a neutral particle with an electric field inside a long non-conducting cylindrical shell of inner radius $r_{a}$ and outer radius $r_{b}$, we show that a geometric quantum phase stems from the missing electric charge per unit length. Thus, we discuss the possibility of existing Aharonov-Bohm-type effects with regard to this geometric quantum phase. Further, we discuss the persistent spin currents.

\end{abstract}

%\pacs{}

\maketitle

\section{Introduction}

The Aharonov-Casher effect \cite{ac} is the appearance of a geometric quantum phase in the wave function of a neutral particle when the magnetic dipole moment of the neutral particle interacts with an electric field produced by a linear distribution of electric charges. The experimental observation of the Aharonov-Casher effect was made by Cimmino {\it et al} \cite{cim} and by Sangster {\it et al} \cite{san,san1} through neutron interferometry. 
We should note here the fact that the effect was first predicted by Heask\'o \cite{extra1}, and the experiments of \cite{extra2} are also worth citing. As discussed by Ionicioiu \cite{ion}, the Aharonov-Casher effect should be considered as the reciprocal effect of the Aharonov-Bohm effect \cite{ab}. It has inspired other works with neutral particles, for instance, its dual effect that was proposed by He and McKellar \cite{mac} and Wilkens \cite{wil} and studies of geometric quantum computation \cite{hol,ion} and quantum holonomies \cite{bf20}. From the perspective of bound states, the Aharonov-Casher effect has been studied in Refs. \cite{bf18,bf21} as an analogue of the Ahaornov-Bohm effect for bound states \cite{pesk}. Another perspective is in studies of persistent spin currents \cite{spin1,spin2,spin3,spin4} and Landau quantization \cite{er,bf19}.

In this work, we discuss the Aharonov-Casher effect \cite{ac} from the interaction of the magnetic dipole moment of a neutral particle with an electric field produced by a uniform distribution of the electric charges inside a long non-conducting cylindrical shell of inner radius $r_{a}$ and outer radius $r_{b}$. This electric field does not fill the region with a cylindrical cavity of radius $r_{a}$. We search for bound states by assuming that there is an impenetrable potential wall located at the inner radius of the cylindrical shell. Then, we discuss the possibility of existing an Aharonov-Bohm-type effect \cite{ab,pesk} associated with a geometric phase that stems from the missing electric charge. Furthermore, we raise a discussion about persistent spin currents \cite{spin1,spin2,spin3,spin4,bf18,bf21}.

The structure of this paper is: in section II, we start by describing the electric field configuration. Then, we show that a geometric quantum phase for a neutral particle with a permanent magnetic dipole moment can arise from the influence of the missing electric charge. Thereby, we discuss an Aharonov-Bohm-type effect for bound states \cite{ab,pesk} associated with the geometric quantum phase yielded by the missing electric charge. We go further by discussing the persistent spin currents; in section III, we present our conclusions.

\section{Missing Aharonov-Casher geometric quantum phase}

We begin by introducing the interaction of the permanent magnetic dipole moment of a neutral particle with an electric field as proposed by Aharonov and Casher \cite{ac} in search of geometric quantum phases. The Schr\"odinger-Pauli equation that describes the quantum description of the interaction of the permanent magnetic dipole moment ($\vec{\mu}=\mu\,\vec{\sigma}$) of a neutral particle with and electric field $\vec{E}$ and a magnetic field $\vec{B}$ is given by \cite{anan1a,ana,anan1c,ac,bf18,bf21} (we shall work with $\hbar=1$ and $c=1$)
\begin{eqnarray}
\mathbb{E}\,\psi=\frac{\hat{\pi}^{2}}{2m}\psi-\frac{\mu^{2}E^{2}}{2m}\,\psi+\frac{\mu}{2m}\left(\vec{\nabla}\cdot\vec{E}\right)\psi+\mu\,\vec{\sigma}\cdot\vec{B}\,\psi.
\label{1.1}
\end{eqnarray}
Observe that $\vec{\sigma}$ corresponds to the Pauli matrices which satisfy the relation $\left(\sigma^{i}\,\sigma^{j}+\sigma^{j}\,\sigma^{i}\right)=2\,\delta^{ij}$. The operator $\hat{\pi}$, in turn, has its components defined as $\hat{\pi}_{k}=-i\,\frac{1}{h_{k}}\,\partial_{k}-\frac{1}{2\,r}\,\sigma^{3}\,\delta_{\varphi\,k}+\mu\,\left(\vec{\sigma}\times\vec{E}\right)_{k}$ \cite{bf21}, the parameter $h_{k}$ corresponds to the scale factors of this coordinate system. In the cylindrical symmetry, the scale factors are $h_{r}=h_{1}=1$, $h_{\varphi}=h_{2}=r$ and $h_{z}=h_{3}=1$ \cite{arf}.

Henceforth, let us consider a long non-conducting cylindrical shell of inner radius $r_{a}$ and outer radius $r_{b}$ ($r_{b}\,>\,r_{a}$), where there is a uniform distribution of the electric charges inside it. This electric charge distribution produces an electric field in the region $r_{a}\,\leq\,r\,\leq r_{b}$ given by (with the units $\hbar=1$ and $c=1$, then, $\epsilon_{0}=1$)
\begin{eqnarray}
\vec{E}=\left[\frac{\rho}{2}\,r-\frac{\rho\,r_{a}^{2}}{2\,r}\right]\,\hat{r},
\label{1.3}
\end{eqnarray} 
where $\rho$ is the uniform volume charge density inside the cylinder and $\hat{r}$ is the unit vector in the radial direction. Due to the absence of the electric field or electric flux inside the region $r\,<\,r_{a}$, the electric field configuration (\ref{1.3}) can be viewed as an analogue of the magnetic quantum dot defined in Refs. \cite{magdot,magdot2,magdot3}, i.e., an electric quantum dot for a neutral particle system, where the neutral particle possesses a permanent magnetic dipole moment.

From Eq. (\ref{1.1}), we have the term $\vec{A}_{\mathrm{AC}}=\mu\,\vec{\sigma}\times\vec{E}$ plays the role of an effective vector potential. This term was introduced by Aharonov and Casher \cite{ac} with the purpose of obtaining a geometric quantum phase for neutral particles, which is known as the Aharonov-Casher geometric quantum phase. Let us assume that the magnetic dipole moment of the neutral particle is aligned along the $z$-axis, then, with the electric field (\ref{1.3}), the effective vector potential $\vec{A}_{\mathrm{AC}}=\mu\,\vec{\sigma}\times\vec{E}$ acquires two contributions:
\begin{eqnarray}
\vec{A}_{\mathrm{AC}1}=\frac{\mu\,\rho\,r}{2}\,\sigma^{3}\,\hat{\varphi},\,\,\,\,\,\vec{A}_{\mathrm{AC}2}=\frac{\mu\,\rho\,r_{a}^{2}}{2\,r}\,\sigma^{3}\,\hat{\varphi}.
\label{1.4}
\end{eqnarray}

The first contribution is given by $\vec{A}_{\mathrm{AC}1}$, which yields an effective uniform magnetic field $\vec{B}_{\mathrm{eff}}=\vec{\nabla}\times\vec{A}_{\mathrm{AC}1}=\mu\,\rho\,\hat{z}$ \cite{er} in the region $r\geq r_{a}$. The second contribution is given by $\vec{A}_{\mathrm{AC}2}$ which yields the geometric quantum phase:
\begin{eqnarray}
\Phi_{\mathrm{MAC}}=\oint\vec{A}_{\mathrm{AC}2}\cdot d\vec{r}= s\,\pi\,\mu\,\rho\,r_{a}^{2},
\label{1.5}
\end{eqnarray}
where the $s=\pm1$ corresponds to the projections of the magnetic dipole moment on the $z$-axis. Note that $\rho\,r_{a}^{2}=\lambda_{\mathrm{M}}$ is the missing electric charge per unit length inside the region $r\,<\,r_{a}$, therefore, the geometric quantum phase (\ref{1.5}) is an analogue of the Aharonov-Casher geometric quantum phase \cite{ac}. Thereby, the geometric quantum phase (\ref{1.5}) can be considered as the inverse of the proposal of the Aharonov-Casher effect \cite{ac}. We call Eq. (\ref{1.5}) the missing Aharonov-Casher geometric quantum phase. It agrees with Ref. \cite{magdot}, where the appearance of the missing magnetic flux is considered as the inverse of the Aharonov-Bohm proposal \cite{ab,pesk}.

Let us substitute the electric field (\ref{1.3}) in Eq. (\ref{1.1}). By using Eqs. (\ref{1.4}) and (\ref{1.5}), the Schr\"odinger equation (\ref{1.1}) becomes  
\begin{eqnarray}
\mathbb{E}\psi&=&-\frac{1}{2m}\left[\frac{\partial^{2}}{\partial r^{2}}+\frac{1}{r}\frac{\partial}{\partial r}+\frac{1}{r^{2}}\frac{\partial^{2}}{\partial\varphi^{2}}+\frac{\partial^{2}}{\partial z^{2}}\right]\psi+\frac{i\,\sigma^{3}}{2m\,r^{2}}\frac{\partial\psi}{\partial\varphi}+\frac{1}{8m\,r^{2}}\,\psi\nonumber\\
&+&\frac{1}{2m\,r^{2}}\left(\frac{\Phi_{\mathrm{MAC}}}{2\pi}\right)\,\sigma^{3}\,\psi+\frac{i}{m\,r^{2}}\left(\frac{\Phi_{\mathrm{MAC}}}{2\pi}\right)\frac{\partial\psi}{\partial\varphi}+\frac{1}{2m\,r^{2}}\left(\frac{\Phi_{\mathrm{MAC}}}{2\pi}\right)^{2}\psi\nonumber\\
&+&\frac{\mu\rho}{4m}\psi-i\frac{\mu\rho}{2m}\,\sigma^{3}\,\frac{\partial\psi}{\partial\varphi}+\frac{\mu^{2}\rho^{2}}{8m}\,r^{2}\psi-\frac{\mu\rho}{2m}\left(\frac{\Phi_{\mathrm{MAC}}}{2\pi}\right)\sigma^{3}\psi.
\label{1.6}
\end{eqnarray}

Observe that the missing Aharonov-Casher geometric phase $\Phi_{\mathrm{MAC}}$ appears in Eq. (\ref{1.6}). This means that the missing Aharonov-Casher geometric phase can influence the interaction of the magnetic dipole moment of the neutral particle with the eletric field in the region $r\,\geq\,r_{a}$. Let us search for the quantum effects associated with the missing Aharonov-Casher geometric phase $\Phi_{\mathrm{MAC}}$. From Eq. (\ref{1.6}), we have $\psi$ is an eigenfunction of the operators $\hat{J}_{z}=-i\partial_{\varphi}$ \cite{schu} and $\hat{p}_{z}=-i\partial_{z}$. Thereby, the solution to Eq. (\ref{1.6}) can be written as
\begin{eqnarray}
\psi\left(r,\,\varphi,\,z\right)=e^{i\left(\ell+\frac{1}{2}\right)\varphi}\,e^{i\,p_{z}\,z}\left(
\begin{array}{c}
f_{+}\left(r\right)\\
f_{-}\left(r\right)\\
\end{array}\right),
\label{1.6a}
\end{eqnarray}
where $\ell=0,\pm1,\pm2,\ldots$ is the eigenvalue of $\hat{L}_{z}$ and $p_{z}$ is the eigenvalue of $\hat{p}_{z}$. Henceforth, we consider $p_{z}=0$, then, we obtain two independent equations for $f_{+}\left(r\right)$ and $f_{-}\left(r\right)$:
\begin{eqnarray}
f''_{s}+\frac{1}{r}\,f'_{s}-\frac{\gamma^{2}}{r^{2}}\,f_{s}-\frac{m^{2}\omega^{2}_{\mathrm{AC}}}{4}\,r^{2}\,f_{s}+\tau\,f_{s}=0,
\label{1.7}
\end{eqnarray}
where $s=+1$ indicates the equation for the function $f_{+}\left(r\right)$, while $s=-1$ indicates the equation for the function $f_{-}\left(r\right)$. Moreover, the parameters $\omega_{\mathrm{AC}}$, $\gamma$ and $\tau$ are defined as follows:
\begin{eqnarray}
\omega_{\mathrm{AC}}&=&\frac{\mu\rho}{m};\nonumber\\
\gamma&=&\ell+\frac{1}{2}\left(1-s\right)-\frac{\Phi_{\mathrm{MAC}}}{2\pi};\label{1.8}\\
\tau&=&2m\mathbb{E}-s\,m\omega_{\mathrm{AC}}\,\gamma-m\,\omega_{\mathrm{AC}}.\nonumber
\end{eqnarray}

We proceed with a change of variables $y=\frac{m\,\omega_{\mathrm{AC}}}{2}\,r^{2}$, then, the radial equation (\ref{1.7}) becomes:
\begin{eqnarray}
y\,f''_{s}+f'_{s}-\frac{\gamma^{2}}{4y}\,f_{s}-\frac{y}{4}\,f_{s}+\frac{\tau}{2m\omega_{\mathrm{AC}}}\,f_{s}=0.
\label{1.9}
\end{eqnarray}

Let us analyse the behaviour of Eq. (\ref{1.9}) as $y\rightarrow\infty$. When $y\rightarrow\infty$, we can write the solution to Eq. (\ref{1.9}) as follows: 
\begin{eqnarray}
f_{s}\left(y\right)=e^{-\frac{y}{2}}\,y^{\left|\gamma\right|/2}\,U\left(\frac{\left|\gamma\right|}{2}+\frac{1}{2}-\frac{\tau}{2m\omega_{\mathrm{AC}}},\,\left|\gamma\right|+1;\,y\right),
\label{1.12}
\end{eqnarray}
where $U\left(a,\,b;\,y\right)$ is the confluent hypergeometric function of the second kind \cite{abra,arf}, whose parameters are defined by $a=\frac{\left|\gamma\right|}{2}+\frac{1}{2}-\frac{\tau}{2m\omega_{\mathrm{AC}}}$ and $b=\left|\gamma\right|+1$. Next, we assume that there exists an impenetrable potential wall at $r=r_{a}$. This gives the boundary condition:
\begin{eqnarray}
f_{s}\left(y_{a}\right)=0,
\label{1.13}
\end{eqnarray}
where $y_{a}=\frac{m\,\omega\,r^{2}_{a}}{2}$. By using (\ref{1.12}), we have from the boundary condition (\ref{1.13}):
\begin{eqnarray}
f_{s}\left(y_{a}\right)\Rightarrow U\left(\frac{\left|\gamma\right|}{2}+\frac{1}{2}-\frac{\tau}{2m\omega_{\mathrm{AC}}},\,\left|\gamma\right|+1;\,y_{a}\right)=0.
\label{1.14}
\end{eqnarray}

Let us first explore the boundary condition (\ref{1.14}) with the case in which $y_{a}$ and $\bar{b}=\left|\gamma\right|+1$ are fixed and $\bar{a}\rightarrow\infty$. In this case, the function $U\left(\bar{a},\,\bar{b};\,y_{a}\right)$ can be written in the form \cite{abra}: 
\begin{eqnarray}
U\left(\bar{a},\,\bar{b};\,y_{a}\right)\propto\cos\left(\sqrt{2\bar{b}y_{a}-4\bar{a}y_{a}}-\frac{\bar{b}\pi}{2}+\bar{a}\pi+\frac{\pi}{4}\right).
\label{1.15}
\end{eqnarray}

Therefore, after substituting Eq. (\ref{1.15}) into Eq. (\ref{1.14}) , we obtain the energy levels:
\begin{eqnarray}
\mathbb{E}_{n,\,\ell,\,s}&=&-\omega_{\mathrm{AC}}\left[n-\frac{s}{2}\left(\ell+\frac{1}{2}\left(1-s\right)-s\frac{\Phi_{\mathrm{MAC}}}{\pi}\right)+\frac{3}{4}\right]\nonumber\\
&-&s\frac{\Phi_{\mathrm{MAC}}}{\pi^{3}}\,\omega_{\mathrm{AC}}\left[1\pm\sqrt{1-\frac{\pi^{3}}{2\,s\,\Phi_{\mathrm{MAC}}}\left(4n+1\right)}\right],
\label{1.16}
\end{eqnarray}
where $n=0,1,2,3,\ldots$ is the radial quantum number.

Hence, from the interaction of the magnetic dipole moment of the neutral particle with the inhomogeneous electric field (\ref{1.3}), we have achieved the discrete spectrum of energy (\ref{1.16}) in the region $r\,\geq\,r_{a}$. The energy levels (\ref{1.16}) are influenced by the missing Aharonov-Casher geometric phase $\Phi_{\mathrm{MAC}}$. Despite the interaction between the magnetic dipole moment of the neutral particle and the effective uniform magnetic field $\vec{B}_{\mathrm{eff}}=\vec{\nabla}\times\vec{A}_{\mathrm{AC}1}=\mu\,\rho\,\hat{z}$ in the region $r\,\geq\,r_{a}$, the energy levels (\ref{1.16}) differ from those of the Landau-Aharonov-Casher levels \cite{er}. The influence $\Phi_{\mathrm{MAC}}$ on the energy levels (\ref{1.16}) is analogous to what is observed in Refs. \cite{magdot,magdot2,magdot3}, where the missing magnetic flux modifies the degeneracy of the Landau levels. In addition, an interesting aspect of the energy levels (\ref{1.16}) is the upper limit of the radial quantum number. This upper limit is given by
\begin{eqnarray}
n_{\mathrm{max}}\,<\,\frac{s\,\Phi_{\mathrm{MAC}}}{2\pi^{3}}-\frac{1}{4}.
\label{1.17}
\end{eqnarray}
Then, $n=0,1,2,3,\ldots,n_{\mathrm{max}}$, otherwise, we would have an imaginary term in the spectrum of energy. Therefore, the upper limit of radial quantum number is determined by the missing Aharonov-Casher geometric phase (\ref{1.5}).

We go further by raising the discussion about the possibility of having an analogue of the persistent spin currents \cite{by,bf18,spin1,spin2,spin3,spin4}. This possibility arises from the fact that the energy levels (\ref{1.16}) depend on the missing Aharonov-Casher geometric quantum phase (\ref{1.5}). From this perspective, based on Refs. \cite{by,spin1,spin2,spin3,spin4}, the persistent spin currents at the temperature is $T=0$ are given by
\begin{eqnarray}
\mathcal{I}=-\sum_{n,\,\ell}\frac{\partial \mathbb{E}_{n,\,\ell,\,s}}{\partial\Phi_{\mathrm{MAC}}}.
\label{1.18}
\end{eqnarray}
Thus, from the energy levels (\ref{1.16}), we obtain
\begin{eqnarray}
\mathcal{I}=\frac{\omega_{\mathrm{AC}}}{2\pi}-s\frac{\omega_{\mathrm{AC}}}{\pi^{3}}\pm\sum_{n}\left[s\frac{\omega_{\mathrm{AC}}}{\pi^{3}}\sqrt{1-\frac{\pi^{3}\left(4n+1\right)}{2\,s\,\Phi_{\mathrm{MAC}}}}+\frac{\omega_{\mathrm{AC}}\left(4n+1\right)}{4\Phi_{\mathrm{MAC}}\sqrt{1-\frac{\pi^{3}}{2\,s\,\Phi_{\mathrm{MAC}}}\left(4n+1\right)}}\right].
\label{1.18a}
\end{eqnarray}

Hence, Eq. (\ref{1.18a}) is an analogue of the persistent spin currents \cite{by,spin1,spin2,spin3,spin4}. It is the expression of the persistent spin currents associated with the missing Aharonov-Casher geometric quantum phase $\Phi_{\mathrm{MAC}}$. Besides, this analogue of the persistent spin currents exists under the same restriction given in Eq. (\ref{1.17}). Otherwise, we would have an imaginary persistent spin current.

Let us return to the boundary condition (\ref{1.14}) and explore another case. Let us consider $y_{a}\ll1$. In this case, the function $U\left(\bar{a},\,\bar{b};\,y_{a}\right)$ can be written in the form \cite{abra,b4}:
\begin{eqnarray}
U\left(\bar{a},\,\bar{b};\,y_{a}\right)&\propto&\frac{\Gamma\left(\bar{b}-\bar{a}\right)}{\Gamma\left(\bar{a}\right)}\,y_{a}^{1-\bar{b}},
\label{1.19}
\end{eqnarray}
where $\Gamma\left(\bar{a}\right)$ and $\Gamma\left(\bar{b}-\bar{a}\right)$ are the Gamma functions. By substituting Eq. (\ref{1.19}) into Eq. (\ref{1.14}), we have that the boundary condition is satisfied only if $\Gamma\left(\bar{a}\right)\rightarrow\infty$. This occurs when $\bar{a}=-n$, where $n=0,1,2,3,\ldots$. In this way, we obtain the energy levels:
\begin{eqnarray}
\mathbb{E}_{n,\,\ell,\,s}=\omega_{\mathrm{AC}}\left[n+\frac{1}{2}\left|\ell+\frac{1}{2}\left(1-s\right)-\frac{\Phi_{\mathrm{MAC}}}{2\pi}\right|+\frac{s}{2}\left(\ell+\frac{1}{2}\left(1-s\right)-\frac{\Phi_{\mathrm{MAC}}}{2\pi}\right)+1\right].
\label{1.20}
\end{eqnarray}
where $n=0,1,2,3,\ldots$ remains the radial quantum number.

Hence, the energy levels (\ref{1.20}) are from the bound states achieved around the cylindrical cavity. They also stem from the interaction of the permanent magnetic dipole moment of the neutral particle with the inhomogeneous electric field (\ref{1.3}). In contrast to the energy levels (\ref{1.16}), there is no upper limit to the radial quantum number. Moreover, when $y_{a}\ll1$, we obtain energy levels which are analogous to the Landau-Aharonov-Casher levels \cite{er} even though an impenetrable potential wall exists at $r=r_{a}$. An interesting aspect of the energy levels (\ref{1.20}) is the influence of the missing Ahaornov-Casher geometric quantum phase $\Phi_{\mathrm{MAC}}$ on them, which does not occur with the Landau-Aharonov-Casher levels \cite{er}. In view of the geometric quantum phase influence, the missing Ahaornov-Casher geometric quantum phase breaks the degeneracy of the Landau-Aharonov-Casher levels \cite{er}.

Furthermore, from dependence of the energy levels (\ref{1.20}) on the missing Aharonov-Casher geometric quantum phase $\Phi_{\mathrm{MAC}}$, the persistent spin currents at the temperature $T=0$ are
\begin{eqnarray}
\mathcal{I}=s\,\frac{\omega_{\mathrm{AC}}}{4\pi}+\sum_{\ell}\frac{\omega_{\mathrm{AC}}}{4\pi}\frac{\gamma}{\left|\gamma\right|}.
\label{1.21}
\end{eqnarray}

Hence, the persistent spin currents (\ref{1.21})) are also influenced by the missing Aharonov-Casher geometric quantum phase $\Phi_{\mathrm{MAC}}$.

Finally, the dependence of the energy levels (\ref{1.16}) and (\ref{1.20}) on the missing Aharonov-Casher geometric phase $\Phi_{\mathrm{MAC}}$ can be considered as an analogue of the Aharonov-Bohm effect for bound states \cite{ab,pesk} or an analogue of the Aharonov-Casher effect for bound states \cite{bf18}. Thereby, we call this quantum effect the Aharonov-Casher effect for missing geometric quantum phase.

\section{Conclusions}

We have made an analogy with the magnetic quantum dot \cite{magdot,magdot2,magdot3} by proposing an electric quantum dot for a neutral particle with a permanent magnetic dipole moment if the neutral particle is confined to the region $r\,<\,r_{a}$, i.e., inside the cylindrical cavity of radius $r_{a}$. In the region $r\,>\,r_{a}$, in turn, we have considered a uniform distribution of the electric charges inside a long non-conducting cylinder. In this region, the uniform distribution of the electric charges has produced the electric field (\ref{1.3}), and thus, we have analysed the interaction of the magnetic dipole moment of a neutral particle with the electric field (\ref{1.3}). We have seen that a geometric quantum phase stems from the missing electric charge per unit length inside the region $r\,<\,r_{a}$. We have shown that this geometric quantum phase is an analogue of the Aharonov-Casher geometric quantum \cite{ac}, thus, we have called it the missing Aharonov-Casher geometric quantum phase $\Phi_{\mathrm{MAC}}$ .

By searching for bound states, we have considered the presence of an impenetrable potential wall at $r=r_{a}$ (in the inner radius of the cylinder). In the first case analysed (fixed $y_{a}$ and $b=\left|\gamma\right|+1$, and $a\rightarrow\infty$), we have obtained a discrete spectrum of energy characterized by its dependence on the missing Aharonov-Casher geometric quantum phase $\Phi_{\mathrm{MAC}}$. Moreover, the radial quantum number possesses an upper limit, which is determined by the missing Aharonov-Casher geometric quantum phase.

In the second case analysed ($y_{a}\ll1$), we have obtained bound states around the cylindrical cavity. The energy levels obtained are analogous to the Landau-Aharonov-Casher levels \cite{er}. We have seen that the missing Ahaornov-Casher geometric quantum phase $\Phi_{\mathrm{MAC}}$ breaks the degeneracy of the Landau-Aharonov-Casher levels, which means that the Landau-Aharonov-Casher levels are modified by $\Phi_{\mathrm{MAC}}$.

In both cases of the interaction of the permanent magnetic dipole moment with the inhomogeneous electric field (\ref{1.3}) analysed, the energy levels depend on the missing Aharonov-Casher geometric quantum phase $\Phi_{\mathrm{MAC}}$. In view of this aspect of the energy levels, we have calculated the persistent spin current associated with $\Phi_{\mathrm{MAC}}$. Besides, this quantum effect can be viewed as an Aharonov-Bohm-type effect \cite{pesk,ab} or an analogue of the Aharonov-Casher effect for bound states \cite{bf18}.

\acknowledgments{The authors would like to thank CNPq for financial support.}

\section*{Data Availability Statement}

%This work does not have any experimental data. 

No data were created or analyzed in this article.

%\section*{Competing interests}

%On behalf of all authors, the corresponding author states that there is no conflict of interest.

%We have no competing interests.

%\section{Ethics statement}

%This research poses no ethical considerations.

\end{document}